\documentclass[aps,prb,twocolumn,notitlepage,citeautoscript,superscriptaddress,footinbib,longbibliography]{revtex4-2}

\usepackage{times}
\usepackage{amsmath,amssymb,mathrsfs,bm,setspace}
\usepackage{dcolumn}
\usepackage{graphicx}
\usepackage{latexsym,hyperref}
\usepackage{xspace}

\usepackage{color}

\usepackage{bm}
\usepackage{epsfig}
\usepackage{amsmath}
\usepackage{color}

\newcommand{\nel}{{n_\mathrm{el}}}

\begin{document}

\title{
	Hund's metallicity enhanced by van Hove singularity in cubic perovskite systems
}

\author{Hyeong~Jun \surname{Lee}}

\affiliation{Center for Theoretical Physics of Complex Systems, Institute for Basic Science (IBS), Daejeon 34126, Republic of Korea}

\author{Choong~H. \surname{Kim}}
\email[]{chkim82@snu.ac.kr}

\affiliation{Center for Correlated Electron Systems, Institute for Basic Science (IBS), Seoul 08826, Republic of Korea}
\affiliation{Department of Physics and Astronomy, Seoul National University, Seoul 08826, Republic of Korea}

\author{Ara \surname{Go}}
\email[]{arago@jnu.ac.kr}

\affiliation{Department of Physics, Chonnam National University, Gwangju 61186, Republic of Korea}
\affiliation{Center for Theoretical Physics of Complex Systems, Institute for Basic Science (IBS), Daejeon 34126, Republic of Korea}

\begin{abstract}
    A van Hove singularity (VHS) often significantly amplifies the electronic instability of a crystalline solid, including correlation-induced phenomena such as Hund's metallicity.
    We perform a systematic study on the interplay between Hund’s coupling and electronic structures with a VHS focusing on Hund’s metallicity. 
    We construct a simplified tight-binding model targeting cubic perovskite materials and test the effects of the VHS utilizing dynamical mean-field theory with an exact diagonalization solver.
    The quasiparticle weight and the low-frequency power exponent of the self-energy provide a quantitative estimation of metallicity over the phase diagram.
    We find the VHS to substantially enhance Hund's metallicity.
    The results here suggest a range of parameters through which a VHS can bring great synergy with Hund's coupling.

\end{abstract}

\date{\today}
\maketitle

\section{Introduction}								

Numerous exotic and intriguing quantum phenomena have been realized in condensed matter systems,
	 some of which arise from correlation effects with electron--electron interactions.
The most well-known example is the Mott metal--insulator transition~\cite{Imada1998}, where the electronic correlations in proximity of the Mott insulator, or so-called Mottness,
    can be captured with an effective single-orbital Hubbard model or trivial atomic multiplet structures.
In addition, it has been revealed that Hund's rule coupling becomes a key factor
	in understanding the complicated correlation effects of materials and their multiorbital nature~\cite{Georges2013}.

Hund's-coupling-driven electronic correlations have been studied extensively within various multiorbital systems~\cite{Georges2013}.
Many studies have shown that Hund's coupling strengthens the effective correlation of the system in a different way than by Hubbard interaction~\cite{Haule2009,Yin2011,Werner2008,Werner2009,DeMedici2011PRL,DeMedici2011PRB}.
For a system with moderate Hubbard interaction $U$, even though the system is far from the Mott insulator phase,
Hund's coupling generates strong band renormalization~\cite{Pruschke2005,Haule2009} and incoherent transport properties with spin freezing~\cite{Werner2008,AJKim2017}.
These phenomena show a strong filling dependence and are maximized in non-singly-occupied and non-half-filled systems
such as Fe-based superconductors~\cite{Haule2009,Yin2011,Werner2012,Liebsch2012,DeMedici2014},
     metallic ruthenates~\cite{Liebsch2000,Pchelkina2007,Mravlje2011,Dang2015,Sutter2019},
     BaOsO$_3$~\cite{Bramberger2021}, and Sr$_2$MoO$_4$~\cite{Karp2020}.
In this respect, such materials began to be classified into a new phase, called the Hund's metals.

Recently, theoretical studies have explored the influence of the van Hove singularity (VHS) on the properties of Hund's metals~\cite{Mravlje2011,HJLee2020,Karp2020,Bramberger2021}.
While the effects of the VHS on various materials have been discussed, systematic theoretical investigations such as the studies of the single-band case in Refs. \cite{Zitko2009,Sebastian2010} have not yet come out with multiorbital models that include Hund's coupling.
This is a difficult subject since the effect of the VHS and other details of the band structure cannot be separated in real materials.
Thermal fluctuations even tend to mask the VHS effects~\cite{Karp2020,HJLee2020}.
Thus, treating the simplest model is necessary for a clear explanation of the effects of the VHS and Hund's coupling in relation to realistic electronic structures.

In this work, we focus on a $t_{2g}$ band model in a cubic lattice by employing dynamical mean-field theory (DMFT) with an exact diagonalization (ED) solver.
We investigate the Hund's metallicity of the model, and in particular, focus on how it is affected by a VHS.
We construct a ground state phase diagram varying Coulomb interaction and electron occupancy, with which we observe how the metal--insulator transition occurs with and without Hund's coupling. We also check the characteristic self-energies and density of states over the phase diagram.
As temperature increases, Hund's metallicity emerges with nonzero Hund's coupling.
We perform systematic calculations for a wide range of parameters to clarify the interplay between Hund's metallicity and the VHS.

\section{Results}							
\label{sec:model}
\subsection{Tight-binding Hamiltonian}				
\begin{figure}[tb]
        \includegraphics[width=0.7\columnwidth]{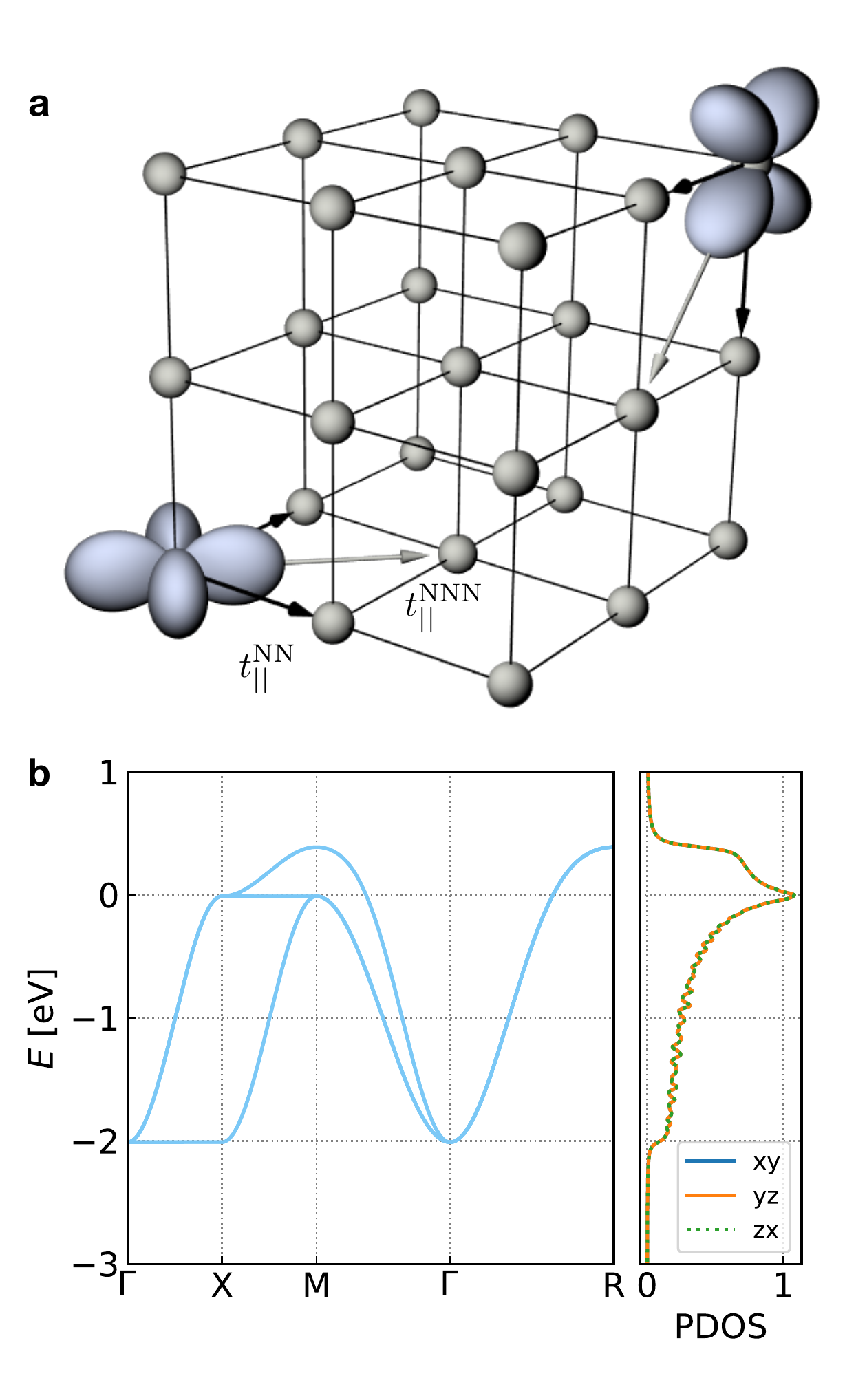}
        \caption{
		Lattice and electronic structure of our model system.
		{\bf a} Schematic illustration of our degenerate three-band system.
		Orbitals are shown with planner hoppings described in Eq.~(\ref{eq:param_hopping}).
		{\bf b} Band dispersion and density of states of non-interacting bands.
	       	The Fermi level lying near the VHS corresponds to $n_{\mathrm{el}}=3.91$.
        }
\label{fig:dosNonint}
\end{figure}

Motivated by our previous work on a ruthenate system~\cite{HJLee2020},
we construct a tight-binding Hamiltonian on a cubic lattice for degenerate $t_{2g}$ bands with a VHS as follows.
On a cubic basis, each orbital has nonzero nearest-neighbour intraorbital hopping amplitudes $t_{\mathrm{NN}}$ along its orbital plane.
We further include next-nearest-neighbour (NNN) hopping $t_{\mathrm{NNN}}$ to introduce particle--hole asymmetry.
The nonzero contribution can be written as
\begin{align}
t^{\alpha}_{(\pm1,0,0)}	& = t_{\mathrm{NN}} \quad \mathrm{for} \quad \alpha \in \{ xy,zx \},		\nonumber \\
t^{\alpha}_{(0,\pm1,0)}	& = t_{\mathrm{NN}} \quad \mathrm{for} \quad \alpha \in \{ xy,yz \},		\nonumber \\
t^{\alpha}_{(0,0,\pm1)}	& = t_{\mathrm{NN}} \quad \mathrm{for} \quad \alpha \in \{ yz,zx \},		\nonumber \\
t^{xy}_{(\pm1,\pm1,0)}	= t^{yz}_{(0,\pm1,\pm1)}	& = t^{zx}_{(\pm1,0,\pm1)}	= t_{\mathrm{NNN}}, \nonumber \\
t^\alpha{(\mathbf{r}_i-\mathbf{r}_j)}		& = 0 \quad \mathrm{otherwise,} 					\; 
\label{eq:param_hopping}
\end{align}
where $t^\alpha\mathbf{r}$ is the hopping amplitude between two $\alpha$-orbitals with displacement $\mathbf{r}$.
This is illustrated in Fig.~\ref{fig:dosNonint}(a).

In this work, we use ($t_{\mathrm{NN}}$, $t_{\mathrm{NNN}}$) = ($-0.3$, $-0.1$) eV unless otherwise specified.
The choice of parameters is based on previous studies on cubic perovskite systems of transition metal oxides such as 
vanadates~\cite{Solovyev2008,Zhong2013}, titanates~\cite{Solovyev2008}, and ruthenates~\cite{Liebsch2003}.
Figure~\ref{fig:dosNonint}(b) shows the band structure and an orbital-projected density of states (DOS) with the parameters above.
The model meets the requirements---three bands and a VHS---to test how a VHS affects Hund's metallic behaviour.

\subsection{Rotationally invariant on-site interaction}	
On top of the $t_{2g}$ tight-binding Hamiltonian of Eq.~(\ref{eq:param_hopping}), 
we introduce rotationally invariant Slater--Kanamori interaction 
\begin{eqnarray}
	\hat{\mathcal{H}}_{\textrm{int}} &=& (U - 3J_H)\frac{\hat{N}(\hat{N}-1)}{2} + \frac{5}{2} J_H \hat{N} 
	\nonumber \\ 
	&& - 2J_H \hat{\bm{S}}^2  - \frac{J_H}{2} \hat{\bm{L}}^2,
	\label{eq:Hint}
\end{eqnarray}
where $U$ and $J_H$ are the Hubbard interaction and Hund's coupling strength, respectively, and $\hat{N}$ and $\hat{\bm{S}}(\hat{\bm{L}})$ denote total number and total spin (orbital) angular momentum operators
defined as $\hat{N}=\sum_\mu \hat{n}_{\mu}, \hat{\bm{S}}=\sum_\mu \hat{\bm{s}}_{\mu}$ and $\hat{\bm{L}}=\sum_\mu \hat{\bm{l}}_{\mu}$. 
We denote $\hat{n}_{\mu}, \hat{\bm{s}}_{\mu}$, and $\hat{\bm{l}}_{\mu}$ as the number, spin, and orbital operators of the $\mu$-th electron in the $t_{2g}$ shell.

The Hubbard interaction dominates the on-site Coulomb repulsion, and may suppress the electron movement between sites when the electron occupancy is an integer, making the system a Mott insulator.
The number of electrons minimizing the energy contribution of the first term in Eq.~(\ref{eq:Hint}) differs for a given chemical potential.
Moreover, the energy levels of multiplets within the same electron number sector further split as Hund's coupling is introduced.


\label{sec:results}

\subsection{Non-interacting dispersion and density of states}						
Figure~\ref{fig:dosNonint}(b) shows the dispersion and density of states projected on the $t_{2g}$ orbitals of our tight-binding model described in Eq.~(\ref{eq:param_hopping}).
The non-interacting DOS acts as a leading indicator that helps us figure out the correlation effects of the interaction between electrons~\cite{Belozerov2018}.
The orbital characters, i.e. bandwidth, VHS locations, etc., as well as the symmetries associated with orbital degrees of freedom can play a significant role in electronic correlations.
Here, three energetically degenerate bands with a bandwidth of about 2.5 eV exhibit a two-dimensional band character, represented by a VHS with a large weight in the middle.
The VHS is shifted from the band center by the existence of NNN hopping. 
When the Fermi level crosses the VHS, this system has $n_{\mathrm{el}}=3.91$ (where $n_{\mathrm{el}}$ is electron concentration).
The electronic structure mimics that of cubic perovskite ($AB$O$_3$) systems of transition metal oxides.

\subsection{Ground state properties}

\begin{figure}[tb]
        \includegraphics[width=\columnwidth]{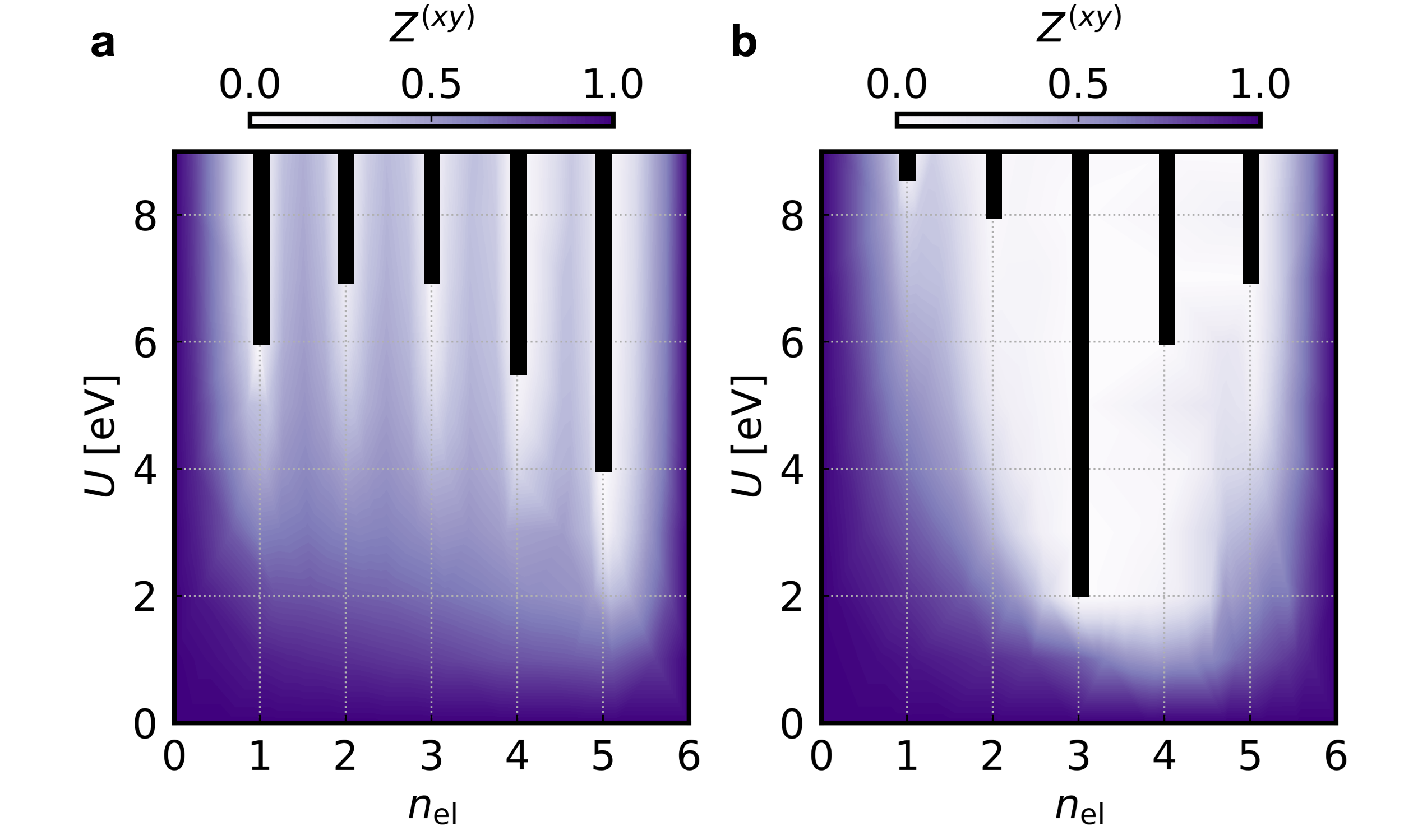}
        \caption{
		Renormalization factor $Z$ in the $U$--$n_{\mathrm{el}}$ plane.
		We estimate and plot $Z$ at $T=0$ K for {\bf a} $J_H=0$ and {\bf b} $J_H=U/6$.
		Mott insulators with integer fillings are marked with black bars.
        }
\label{fig:Zcubic}
\end{figure}

The ground state of this model is either a Mott insulator or a Fermi liquid depending on the parameters.
To investigate the genuine effects of the VHS on interaction-induced correlations,
we first solve the self-consistent DMFT equation at $T=0$ excluding thermal fluctuations.
By analyzing the ground state self-energy for various values of interaction parameters and electronic concentration, 
we observe how the VHS affects correlation effects.

\textit{Phase diagram and Hund's coupling.}				
The Mott insulating phase can be identified by a diverging self-energy at the Fermi level.
Sufficiently large mutual Coulomb interaction can induce the divergence, opening a gap at integer filling.
One can estimate the critical strength of $U_c$ that stabilizes the Mott phase based on the atomic limit with Eq.~(\ref{eq:Hint})~\cite{Rozenberg1997,Florens2002,Ono2003}.
The Mott gap of an isolated atom itself is $\nel$-independent;
once embedded in a lattice though, the kinetic energy gain from multiplet degeneracy competes with the interaction to open a gap~\cite{Florens2002,Ono2003}.
When $J_H=0$, the degeneracy is maximal at half-filling and minimal at $\nel=1$ or $5$.
This means that the system requires larger $U$ to open a gap for half-filling.
We note that the local interaction in Eq.~(\ref{eq:Hint}) is particle--hole symmetric with respect to the half-filling line ($\nel=3$).
If the underlying noninteracting DOS is particle--hole symmetric, then the critical interaction strength $U_c$ where a metal--insulator transition occurs would also be symmetric.
Accordingly, the maximal $U_c$ could be obtained at $\nel=3$ as shown in earlier works~\cite{Ono2003}.

In our model, the NNN hopping terms introduce a particle--hole asymmetric DOS as well as a VHS close to the Fermi level at $\nel\sim 4$.
The black bars in Fig.~\ref{fig:Zcubic} mark the Mott insulating regime.
The strongly asymmetric DOS accompanying the VHS induces a sharp contrast for $\nel < 3$ and $\nel > 3$.
Overall, $U_c(\nel)$ is smaller for $\nel>3$, and the maximal $U_c(n_\mathrm{el})$ occurs at $\nel=2$ instead of $\nel=3$.
This is due to a suppression of the dispersion around $\nel\sim 4$, which reduces kinetic energy gain.

The change in critical interaction strength by Hund's coupling is also highly $\nel$-dependent in our model~\cite{Lombardo2005,Werner2008,Werner2009,DeMedici2011PRL,DeMedici2011PRB}.
The half-filled case ($\nel=3$) is easily gapped with relatively small $U_c$ by forming a spin moment supported by the Hund's coupling.
The $U_c$ for other integer fillings becomes larger, but the asymmetry with respect to $\nel=3$ persists.


\textit{Renormalization factors at $T=\mathit{0}$.}				
The effect of the VHS is clearly captured in the renormalization factor $Z$, which colors the phase diagram in Fig.~\ref{fig:Zcubic}.
In the Fermi liquid regime, the imaginary part of the diagonal self-energies follows the relation
$\lim_{\omega \rightarrow 0} \mathrm{Im} \Sigma_{\alpha \alpha} (i\omega) \sim  (1-Z_\alpha^{-1})\omega$,
where $Z_\alpha$ is the renormalization factor of the $\alpha$-th component in $t_{2g}$.
When the system is a Mott insulator, $Z$ becomes zero according to the above relation by the diverging self-energy at $\omega \rightarrow 0$.
It may appear reasonable to have $Z=0$ for an insulating phase; however, in principle, the relation applies only to Fermi liquids.
Hence, we mark the Mott phase with black bars in Fig.~\ref{fig:Zcubic}.

All the metallic states in our phase diagram at $T=0$ are Fermi liquid.
We extract $Z_{xy}$ ($=Z_{yz}=Z_{zx}$) by fitting the self-energy on the Matsubara frequencies.
Without Hund's coupling [Fig.~\ref{fig:Zcubic}(a)], the small white $Z$ regions are focused on the integer fillings.
As $U_c$ is the smallest at $\nel=5$, the bad metallic region extends to lower $U$ at the same filling.

Hund's coupling reinforces the bad metal with an exception at $\nel=3$.
In the half-filled case, a metal-to-insulator transition occurs at smaller $U_c$, and a reduction of $Z$ is shown in a narrower range of $U$.
However, a substantial suppression of $Z$ is found at $\nel = $ 2 and 4, where we expect the Janus-faced role of Hund's coupling~\cite{DeMedici2011PRL}.
The asymmetric behaviour (the difference between $\nel=2$ and $\nel=4$) is also enhanced by the Hund's coupling in Fig.~\ref{fig:Zcubic}(b).
While $U_c$ is the lowest at $\nel=3$, the bright region is enlarged near $U=0$ at $\nel=4$, implying a synergy between the Hund's coupling and the VHS.

\begin{figure}[tb]
        \includegraphics[width=\columnwidth]{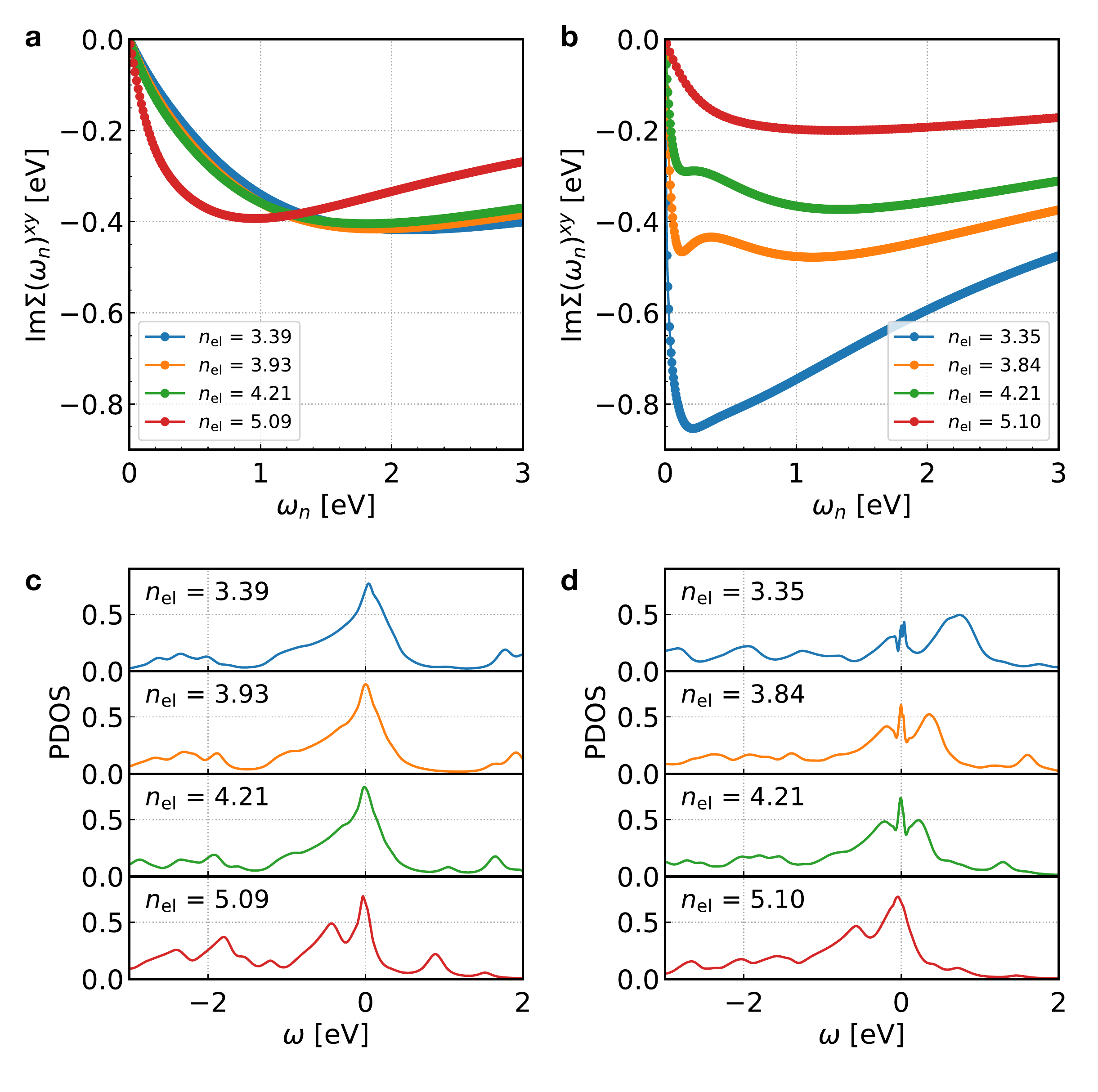}
        \caption{
		Comparison of single-particle spectra.
		{\bf a, b} Imaginary part of the self-energy and {\bf c, d} density of states
			for (left column) $J_H=0$ and (right column) $J_H=1/3$ eV with various electron occupations at $T=0$.
		We use a broadening scheme that depends on frequency in the real-frequency spectra hereafter.
        }
\label{fig:selfenergycubic}
\end{figure}

\textit{Single-particle spectra with Hund's coupling.}					
The renormalization factor discussed so far is now evaluated from the zero-frequency limit of the self-energy.
We inspect the correlations effects through the self-energy at larger frequency values in a systematic way, from which we find that the self-energy reveals structural changes depending on the interaction parameters and electron concentration.
The Matsubara self-energies of our results exhibit two distinct contributions: one from the Hubbard interaction, and one from Hund's coupling.
It has been commonly observed that a metallic self-energy has a single hump for $\omega_n>0$ when the Hubbard interaction is the dominant energy scale, as shown in Fig.~\ref{fig:selfenergycubic}(a).
The hump approaches the Fermi level as $U$ increases. Once the system becomes a Mott insulator, the self-energy diverges at the Fermi level.
Hund's coupling can develop a separated peak near the Fermi level in the self-energy, a behaviour that is prominent when $\nel$ is close to the fillings ($\nel = $ 2 or 4) where the Janus-faced role of the Hund's coupling emerges.

Figure~\ref{fig:selfenergycubic}(a) and \ref{fig:selfenergycubic}(b) show the imaginary part of the Matsubara self-energies with $U=2$ eV for $J_H=0$ and $J_H=U/6$, respectively.
As we discussed above, the hump is closer to the Fermi level around $\nel=5$ in Fig.~\ref{fig:selfenergycubic}(a).
The Mott transition occurs at relatively smaller $U$ at $\nel=5$, and therefore $U=2$ is a shorter distance away from the transition point, as illustrated in Fig.~\ref{fig:Zcubic}(a).
Hund's coupling brings an additional structure near the Fermi level in the self-energy.
In Fig.~\ref{fig:selfenergycubic}(b), the imaginary part of the self-energy shows a sharp peak close to the Fermi level, when $\nel$ is around 4.
This is accompanied by a significantly reduced $Z$, as the slope of the self-energy is steep in the limit of $\omega_n \rightarrow 0$.

The corresponding spectral functions in Fig.~\ref{fig:selfenergycubic}(c) and \ref{fig:selfenergycubic}(d) reflect the structural change of the Matsubara self-energy.
In both Fig.~\ref{fig:selfenergycubic}(c) and \ref{fig:selfenergycubic}(d), we find an obvious $\nel$-dependence of the spectral functions as an overall shift along the real-frequency axis.
As $\nel$ increases, the spectral functions move left, pushing more weights below the Fermi level.
Additionally, Hund's coupling yields a distinctive feature in the spectral functions when the Matsubara self-energy has another clear peak adjacent to the Fermi level.
More specifically, it develops a sharp coherent peak at the Fermi level sandwiched by two dips, as represented in the upper three panels of Fig.~\ref{fig:selfenergycubic}(d).

\subsection{Hund's physics at $T>0$}					

\begin{figure}[bt]
        \includegraphics[width=\columnwidth]{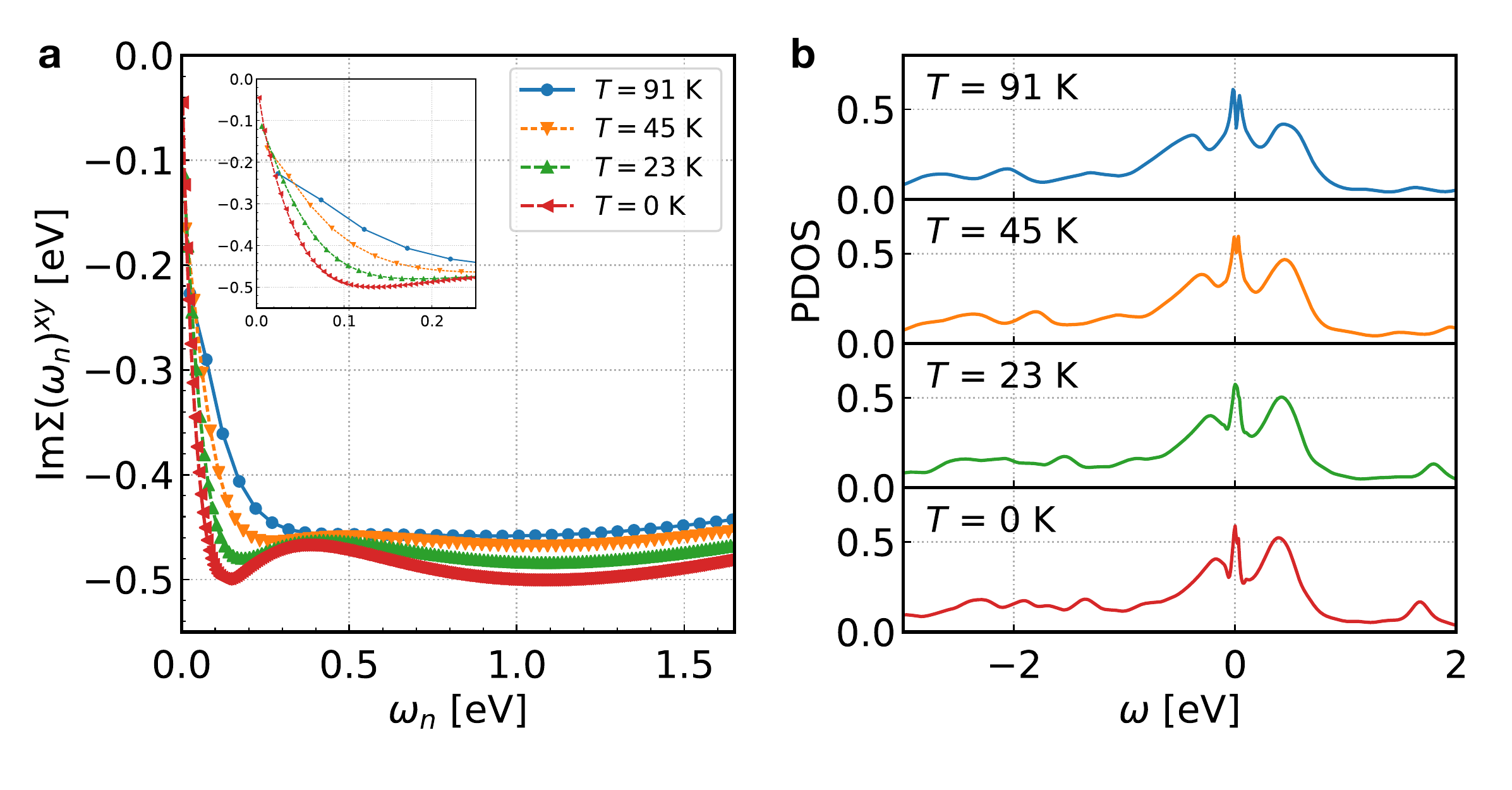}
        \caption{
		Single-particle spectra at $T \geq 0 $ K.
		Data for ($U$, $J_H$) = (2, $\frac{1}{3}$) eV around $n_{\mathrm{el}}\simeq3.8$ with different temperatures:
		{\bf a} imaginary part of the self-energy at imaginary frequencies, and
		{\bf b} density of states at real frequencies.
		The inset shows the data of (a) in a narrow frequency range, $\omega_n<0.3$ eV.
        }
\label{fig:Tcubic}
\end{figure}

\begin{figure}[bt]
        \includegraphics[width=\columnwidth]{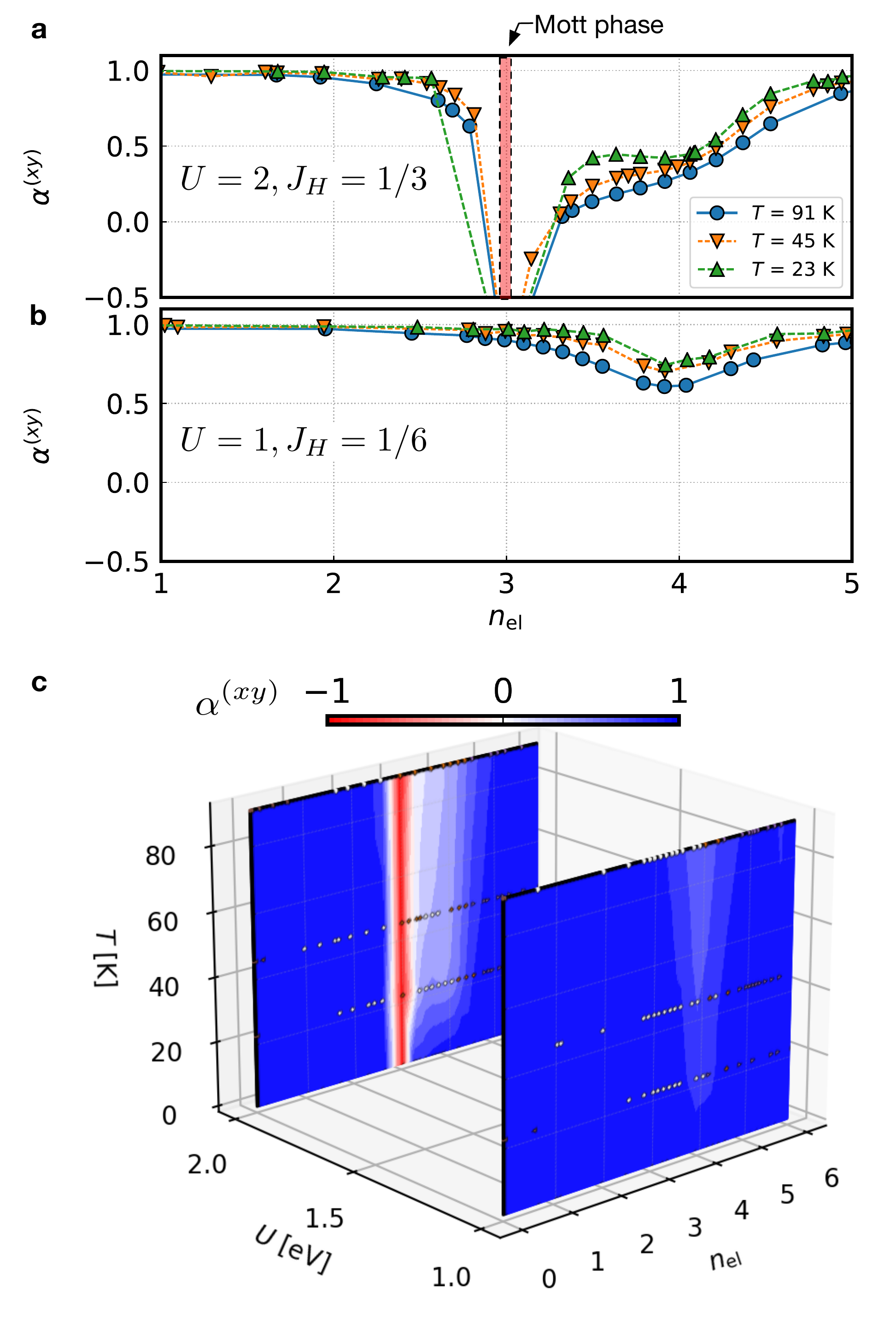}
        \caption{
		Power exponent $\alpha$ of the self-energy. We estimate (see main text) and plot $\alpha$ as a function of electron occupancy $n_{el}$ 
			for {\bf a} ($U$, $J_H$) = (2, $\frac{1}{3}$) eV and {\bf b} (1, $\frac{1}{6}$) eV.
		{\bf c} The same data in the $T$-$n$ plane, in which we also include the $T=0$ data and interpolate them.
        }
\label{fig:5}
\end{figure}

\textit{Self-energy and spectral functions for $T>0$.}					
We now analyse how thermal fluctuations interplay with Hund's coupling in our degenerate three-band model. 
To observe the effects of thermal fluctuations, we focus on the ($U$, $J_H$) = (2, $\frac{1}{3}$) eV region around $n_{\mathrm{el}}=4$.
In this region, the system is strongly incoherent with small values of $Z$, induced by Hund's coupling, where the VHS is sufficiently close to the Fermi level.
We plot the self-energies for various values of temperature in Fig.~\ref{fig:Tcubic}(a).
At $T=0$, there is a peak at $\omega\simeq0.2$ eV and a broad hump around $\omega \sim 1.1$ eV.
As the temperature increases, these features move toward each other with a concurrent reduction in the overall amplitude of $\mathrm{Im}\Sigma(i\omega_n)$.

Meanwhile, the low-frequency part ($\omega_n<0.1$ eV) of the self-energy at $T\lesssim 23$ K shows a linear frequency dependence.
We note that this linearity in the Matsubara frequency persists at nonzero temperatures as long as the system is a Fermi liquid~\cite{Werner2008,Liebsch2010,Werner2012,HJLee2020}. 
Above 23 K, the linearity breaks down, and the Fermi liquid description can no longer be applied to the self-energy.
The deviation from linearity starts to noticeably grow when we further increase temperature above 45 K.
The largest deviation is found at the lowest Matsubara frequency $\omega_0$ [see the inset of Fig.~\ref{fig:Tcubic}(a)].

Figure~\ref{fig:Tcubic}(b) shows the corresponding single-particle spectral functions.
All the spectral functions for $T>0$ exhibit a sharp coherent peak, as they do at $T=0$.
As $T$ increases, the width of the coherent peak grows, which corresponds to the peak of the Matsubara self-energy moving away from the Fermi level.
A breakdown of the Fermi liquid is also observed in the spectral functions.
The data with the two highest temperatures in Fig.~\ref{fig:Tcubic}(b) have a pseudogap at the Fermi level, which is a consequence of the nonzero imaginary part of the self-energies as $\omega_n\rightarrow 0$.
This is consistent with previous reports on other multiorbital systems~\cite{Pruschke2005,Liebsch2011,Liebsch2012,Werner2016}.
The Hund's-driven correlations at $T>0$ enhance this pseudogap behaviour.

\textit{Power exponent of Matsubara self-energy.}					
To quantify the correlation-induced incoherence, we extract a power exponent from the frequency-dependence of the Matsubara self-energy.
We use the two lowest Matsubara frequencies to compute the power exponent, 
\begin{align}
\alpha \approx \frac{\mathrm{log}| \mathrm{Im}\Sigma(i\omega_0)| - \mathrm{log}| \mathrm{Im}\Sigma(i\omega_1) | }{ \mathrm{log}| \omega_0 | - \mathrm{log}| \omega_1|  },
    \label{eq.alpha}
\end{align}
assuming
$\mathrm{Im}\Sigma(i\omega_n) \sim (\omega_n)^\alpha$~\cite{Werner2008,Liebsch2010,Werner2012,Werner2016,HJLee2020}.
In the Fermi liquid phase, we obtain $\alpha=1$.
Then from the deviation from $\alpha=1$, we acquire a quantitative estimation of how far the system is from the Fermi liquid.
For metals, $\alpha$ is between 0 and 1, while for the Mott insulating case, $\alpha=-1$ because of the diverging self-energy.

We plot $\alpha$ for various temperatures with ($U$,~$J_H$) =~(2,~$\frac{1}{3}$) eV in Fig.~\ref{fig:5}(a).
Since all metallic phases are Fermi liquids in the ground state ($T=0$), $\alpha$ is fixed at 1.
For $T>0$, $\alpha$ presents a very asymmetric distribution about the half-filled line.
If the electron occupancy is less than 2, the linearity of the self-energy ($\alpha=1$) is maintained up to about 90 K.
As $\nel$ approaches half-filling from $\nel<3$, $\alpha$ begins to decrease, albeit more slowly compared to the $\nel>3$ region.
This indicates that the Hund's metal, a crossover phase between a Fermi liquid and Mott insulator, is stabilized at lower temperatures by the VHS.
Traces of the VHS are observed with a downward-facing hump around $\nel=4$ as $T$ decreases.
However, the existence of the Mott phase at $\nel=3$ also leads to a rapid change in $\alpha$, making it difficult to distinguish the effects of the VHS through $\alpha$.

To differentiate and observe only the VHS effects, we further reduce $U$.
In Fig.~\ref{fig:5}, we display two data sets, ($U, J_H$) = (2, $\frac{1}{3}$) and ($U, J_H$) = (1, $\frac{1}{6}$), for comparison.
We extract their $\alpha$ for various electronic concentrations and interpolate the $\alpha$ values in Fig.~\ref{fig:5}(c).
For ($U, J_H$) = (1, 1/6), the system is metallic over the entire range of parameters we use in Fig.~\ref{fig:5}(b).
Excluding the effects of the Mott phase, the regime where $\alpha$ deviates from 1 is concentrated around the VHS.
This implies that the hump around $\nel=4$ is a continuation of the effects induced by the VHS.

\begin{figure}[bt]
        \includegraphics[width=0.9\columnwidth]{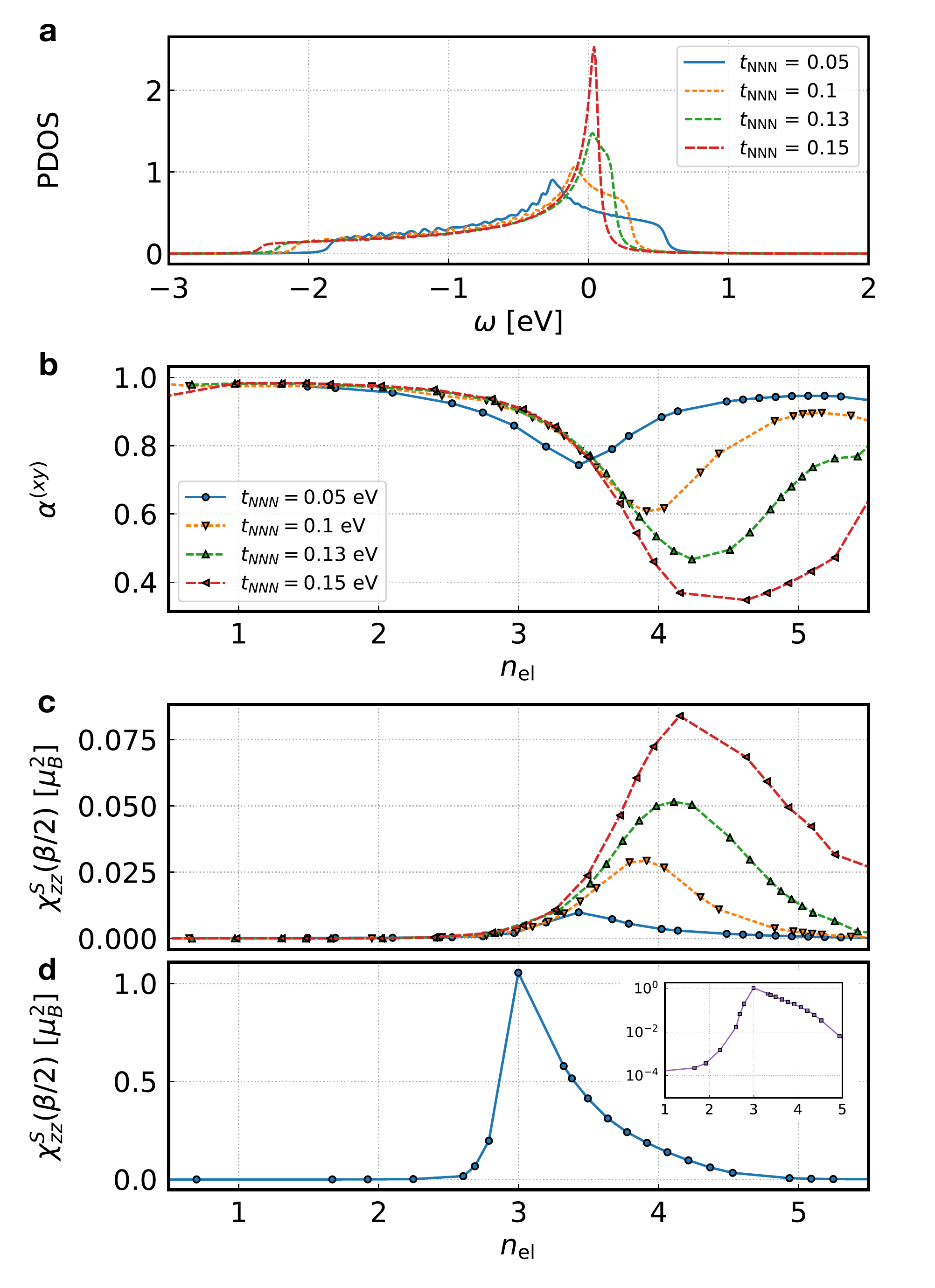}
        \caption{
		Effects of the VHS for various $t_{\mathrm{NNN}}$ values.
		{\bf a} Density of states, {\bf b} power exponent, and {\bf c} long-time spin correlator $\langle \hat{S}_z(\beta/2) \hat{S}_z \rangle$
			for different values of the NNN hopping parameter of Eq.~(\ref{eq:param_hopping}): $t_{\mathrm{NNN}} = $ 0.05, 0.1, 0.13, and 0.15, with ($U, J_H$) = (1, $\frac{1}{6}$).
		{\bf d} Long-time spin correlator for ($U, J_H$) = (2, $\frac{1}{3}$) with $t_{\mathrm{NNN}} = $ 0.1.
        }
\label{fig:tNNN}
\end{figure}

\textit{Pinning Hund's metal by VHS.}					
According to the above results, the Hund's metal appears to be strengthened near the VHS.
To verify this, we tune our tight-binding model with various values of the NNN hopping parameter $t_{\mathrm{NNN}}$ of Eq.~(\ref{eq:param_hopping}).
We choose $t_{\mathrm{NNN}}$ as 0.01, 0.05, 0.13, and 0.15 eV.
The resulting DOS thus has a VHS at different energies, as shown in Fig.~\ref{fig:tNNN}(a).
As the location of the VHS moves to a higher/lower energy in the DOS, the corresponding Fermi level lying in the VHS yields an increase/decrease of $\nel$.
We obtain $\nel \simeq $ 3.4, 3.9, 4.6, and 4.8 for $t_{\mathrm{NNN}} = $ 0.01, 0.05, 0.13, and 0.15 eV, respectively.
Their bandwidths remain almost the same, while $t_{\mathrm{NNN}}$ affects the weight of the VHS.

We also performed the DMFT calculation for these tight-binding Hamiltonians.
Since the effects of the VHS are prominent with thermal fluctuations,
we focus on $T=91$ K for ($U$, $J_H$) = (1, $\frac{1}{6}$) to test the Hund's metallicity systematically.
Figure~\ref{fig:tNNN}(b) illustrates the power exponent $\alpha$ extracted by Eq.~(\ref{eq.alpha}).
The overall crossover can either be weakened or enhanced according to the weight of the DOS at the VHS.
The decrease of $\alpha$ is maximal around $\nel$, where the VHS is located at the Fermi level.
There is, however, a deviation between the location of the minimal value and the Lifshitz transition point for $t_{\mathrm{NNN}} \geq 0.013$.
This may be attributed to the high level of thermal fluctuations.
We confirm that the deviation becomes smaller as we lower the temperature.

The crossover from Fermi liquid to Hund's metal from strong Hund-driven correlations is also studied with the magnetic responses of the system.
A qualitative change of the self-energy from the Fermi liquid is known to appear together with a long-lived local spin moment under time-reversal symmetry~\cite{Werner2008}.
We calculate and analyse a long-time spin correlator defined as $\chi^S_{zz}(\beta/2) = \langle \hat{S}_z(\tau=\beta/2) \hat{S}_z \rangle$.
There is a perfect screening of spin and orbital moments in all metallic states at $T=0$, but
we observe a growth of the long-time correlator at $T>0$.
It is pinned and consistent with the Fermi level on the VHS and the minimal value of $\alpha$, as shown in Fig.~\ref{fig:tNNN}(c).
We also display the long-time correlator for ($U$, $J_H$) = (2, $\frac{1}{3}$), as shown in Fig.~\ref{fig:tNNN}(d).
This long-time correlator shows a monotonic increase toward the Mott state  [Fig.~\ref{fig:Zcubic}(b)] of $n_{\mathrm{el}}=3$.
Although it may appear difficult to distinguish whether the increase comes from the Mott-driven or the Hund-driven correlations,
	 there can be seen a small hump around $\nel=4$ in log-log scale [see the inset of Fig.~\ref{fig:tNNN}(d)], a VHS effect.

\section{Discussion}						
\label{sec:conclusions}
In this work, we presented a DMFT-ED study of a realistic $t_{2g}$ tight-binding model that describes cubic perovskite transition metal compounds.
In wide ranges of electronic concentration and temperature reaching $T=0$,
we have investigated the effects of Hund's coupling, the roles of the VHS, and the interplay between them through single-particle and magnetic responses.

All the metallic phases at $T=0$ are Fermi liquids. As $U$ increases, the system shows metal--insulator transitions at integer fillings.
Hund's coupling decreases the critical interaction strength $U_c$ for the half-filled case $\nel=3$, while it increases $U_c$ for other fillings.
The asymmetric DOS by next-nearest-neighbour hopping reduces $U_c$ for $\nel>3$ regardless of Hund's coupling.
The metallicity from the Fermi liquid is also suppressed by Hund's coupling, and this suppression is reinforced around the filling where the VHS is at the Fermi level.

Thermal fluctuations further enhance the Hund's-coupling-driven incoherence around the VHS.
We tested this effect by shifting the VHS to different Fermi levels, and found that the Hund's metallic phase extends to lower temperatures around the VHS with the aid of Hund's coupling.

\section{Method}							
\label{sec:method}
\subsection{DMFT with ED impurity solver}				
We use DMFT~\cite{Georges1996} to examine the correlation effects of the rotationally invariant interaction of Eq.~(\ref{eq:Hint}) in our tight-binding Hamiltonian.
DMFT utilizes an effective impurity model, where an impurity is embedded in an electronic bath.
Assuming that the self-energy of the lattice is local, we solve the self-consistency equation and
use the ED method~\cite{Caffarel1994} to solve the impurity model.
The ED method requires bath discretization and fitting the bath energies and hybridizations in order to describe the lattice system.
In detail, we use three (six when considering spin) correlated and nine (18) bath orbitals,
	which satisfy the minimal requirement~\cite{Koch2008,Senechal2010,Liebsch2012,Go2015,Fertitta2018} to capture correlation effects in multiorbital physics.

We note that correlated metals at very low temperature can be successfully described
	when the bath orbitals near the Fermi level are given priority out of the finite number of bath orbitals~\cite{Liebsch2012,Linden2020}.
We impose the paramagnetic constraint, and all the results presented in this work preserve time-reversal symmetry.

\section{Acknowledgements}						
This work was supported by the Institute for Basic Science under grants No. IBS-R024-D1 (H.J.L.) and IBS-R009-D1 (C.H.K.).
A.G. is supported by the National Research Foundation of Korea (NRF) under grant No.~NRF-2021R1C1C1010429.

\bibliography{3bands_cubic}

\end{document}